# Document clustering with evolved multi-word search queries

Laurence Hirsch[1] · Robin Hirsch[2] · Bayode Ogunleye[3]

**Abstract**
Text clustering holds significant value across various domains due to its ability to identify patterns and group related information. Current approaches which rely heavily on a computed similarity measure between documents are often limited in accuracy and interpretability. We present a novel approach to the problem based on a set of evolved search queries. Clusters are formed as the set of documents matched by a single search query in the set of queries. The queries are optimized to maximize the number of documents returned and to minimize the overlap between clusters (documents returned by more than one query). Where queries contain more than one word they are interpreted disjunctively. We have found it useful to assign one word to be the root and constrain the query construction such that the set of documents returned by any additional query words intersect with the set returned by the root word. Not all documents in a collection are returned by any of the search queries in a set, so once the search query evolution is completed a second stage is performed whereby a KNN algorithm is applied to assign all unassigned documents to their nearest cluster. We describe the method and present results using 8 text datasets comparing effectiveness with well-known existing algorithms. We note that as well as achieving the highest accuracy on these datasets the search query format provides the qualitative benefits of being interpretable and modifiable whilst providing a causal explanation of cluster construction.

**Keywords**  Document clustering · Search query · Genetic algorithm · Machine learning · Apache Lucene

## 1 Introduction

Clustering algorithms group a collection of documents into subsets or clusters to enable users to explore, organise, summarise, curate and visualise large volumes of text. Documents within a cluster should be similar to each other (cohesion) whilst documents in different clusters should be dissimilar (separation). Text clustering is a central component of text mining and plays a crucial role in enhancing search engine performance and user experience, for example in organising search results, providing personalised search and organising document indexes. Text clustering can be used to analyse search queries and group them based on the intended user meaning. This helps search engines understand the underlying context and nuances of the query, leading to more relevant results that match the user's true information needs [1].

For automated clustering, documents are traditionally represented by a multi-dimensional feature vector where each dimension corresponds to a weighted value of a term within the document collection [2]. Various similarity or distance measures have been proposed and are a central component of text clustering algorithms. Using such a method it is often difficult for a human to understand how the clustering is performed and there has been some criticism of the black box nature of many successful machine learning models, particularly where large datasets may contain human biases and prejudices [3, 4]. As a result of the risks associated with relying on sophisticated machine learning text clustering models which are not completely comprehensible, significant efforts have been made to create explainable systems. Work has been done on alternative models which recognise word order such as using lexical chains to preserve the

✉ Laurence Hirsch
  l.hirsch@shu.ac.uk

  Robin Hirsch
  r.hirsch@ucl.ac.uk

  Bayode Ogunleye
  b.ogunleye@brighton.ac.uk

1 Sheffield Hallam University, Sheffield, UK
2 University College London, London, UK
3 University of Brighton, Brighton, UK

Springer

semantic relationships between words, for example by using WordNet [5, 6]. Efforts have also been made to generate a set of human interpretable rules from 'black box' systems such as support vector machines as in [7]. Word embeddings have been used to improve the accuracy of clustering together with a system to learn interpretable labels [8]. Another effective approach has involved combining neural network models with symbolic representations [9]. Hierarchical clustering produces a dendrogram, a tree-like diagram that visually depicts the clustering process and cluster relationships and has the advantage or not requiring the exact number of clusters to be specified in advance [10].

## 1.1 Motivation

When constructing a search query, a human user will normally try to find a word or combination of words that returns the documents they are interested in but will not return other documents. In their seminal work Manning et al. assert that the *'fundamental assumption'* of information retrieval is what they term the 'cluster hypothesis':

> *Documents in the same cluster behave similarly with respect to relevance to information needs. The hypothesis states that if there is a document from a cluster that is relevant to a search request, then it is likely that other documents from the same cluster are also relevant.*[1]*"* [11]

Following this hypothesis, we have developed a system called eSQ (evolved Search Queries), which uses a Genetic Algorithm (GA) for evolving a set of search queries where a cluster is the documents returned by a single query in the set. The overall objective is to develop an effective clustering system with a natural fit for information retrieval needs and with the following desirable characteristics:

1. Easily interpreted by a human.
2. Modifiable by a human.
3. Provide a causal explanation of cluster construction.

## 1.2 Contribution

We believe this is the first attempt to produce effective document clustering based on simple, explainable multi-word search queries. Unlike other clustering algorithms the query-based algorithm does not rely on distance metrics in its initial phase. The techniques presented should be useful in both text mining, information retrieval and the development of new methods of clustering. The simple disjunctive search queries produced by the eSQ system are accurate, easy to understand, highly scalable and are potentially modifiable by a human analyst. The eSQ system presented here also has a novel fitness test based entirely on the count of unique query hits.

## 2 Background

### 2.1 Common algorithms

Many algorithms have been proposed to achieve document clustering but the most popular is the k-means algorithm. K-means begins with *k* arbitrary centres, typically chosen uniformly at random from the data points. Each point is then assigned to the nearest centre, and each centre is recomputed as the centre of mass of all points assigned to it. These two steps are repeated until the process stabilizes. K-means + + is an enhanced version of the k-means algorithm and uses a randomised seeding technique which is a specific way of selecting initial centres [12, 13]. K-means (and its variants) is still widely used and an active topic of current research [14]. The main disadvantage of k-means is that we must provide the desired number of clusters to the algorithm in advance. Furthermore, clustering algorithms such as k-means are limited in their ability to capture contextual information and semantically explain the reasoning behind the clustering results. Attempts have been made to determine the optimal value of k using genetic algorithms [15].

Agglomerative clustering is a popular alternative to k-means which can determine a suitable value for *k*. Agglomerative clustering is the bottom-up form of hierarchical clustering which treats each data point or object as cluster at the initial stage [16]. In subsequent iterations, using a linkage function the cluster joins the nearest cluster at distance D. The distance D is calculated using metrics such as Euclidean distance. The iteration continues until a large cluster with all the objects is formed [16]. There have been attempts to combine agglomerative clustering and k-means [17].

One of the most commonly used method to represent textual data is term frequency inverse document frequency (TFIDF). However, TFIDF cannot consider the position and context of a word in a sentence. Word embeddings which incorporate the position and context of a word in a sentence have been used to improve the accuracy of clustering results [18].

### 2.2 Clustering with GA

The system we propose uses a GA which is a stochastic global optimisation technique which mimics the process of Darwinian evolution whereby the search for solutions is guided by the principles of selection and heredity [19]. GAs have proved to be an effective computational method, especially in situations where the search space is un-characterized (mathematically), not fully understood, or/and highly dimensional [20]. Clustering a large volume of text is an example of unsupervised problems that fits all these characteristics and GAs have been used here, often as a means of optimizing the allocation of cluster centres [21–23]. Frequently, each chromosome represents a combination of centres which represent the candidate solution to the clustering problem [14]. A ratio of the intra-cluster distance of clusters against the inter-cluster distances between the cluster is often used. Document clustering using a fitness function based on the concept of nearest neighbour separation has also been proposed [24].

---
[1] Our emphasis.

GAs have also been in use for some time to generate rules for text classification [25–27] and clustering [28, 29], which have the advantage of being explainable.

## 2.3 Topic modelling

The system we propose has some similarities to topic modelling which aims to automatically discover latent topics from a collection of documents [30]. Topic modelling and text clustering are both unsupervised machine learning techniques used to analyse, organize and understand large collections of text data. Topics of documents can be found by searching for groups of words that frequently occur together in documents across the collection or by using semantic information in the documents [31–33]. The query words produced automatically by eSQ can be used as topic words. The system differs in that it requires no prior semantic information.

## 3 Materials and methods

### 3.1 Document collections

Different clustering algorithms can produce divergent results when compared to each other on different datasets with different types of text. We, therefore, ran our experiments on 8 different datasets selected from 3 document collections containing very different types of document. Each dataset is labelled in bold.

#### 3.1.1 CrisisLex

An increasing number of short texts are being generated and it has been noted that this environment is complicated by sparsity and high-dimensionality, meaning that the vector space model and normal text clustering methods may not work well [34, 35]. CrisisLex.org is a repository of crisis-related social media data and tools [36]. The 'CrisisLexT6' collection2 contains tweets collected in 2012–13 in different crisis situations. We use 1000 of the tweets from each of the categories. **Crisis3** is created from: Colorado wildfires, Boston bombings and Queensland floods. **Crisis4** is created from Colorado wildfires, Boston bombings, Queensland floods and LA airport shootings.

#### 3.1.2 Newsgroups

In the 20 Newsgroups collection [37] documents are messages posted to Usenet newsgroups, and the categories are the newsgroups themselves. The data on this set is considered particularly noisy and as might be expected does include complications such as duplicate entries and cross postings. We create three datasets from this collection by randomly selecting 400 documents from each of the categories. **NG3** is created from: rec.sport.hockey, sci.space and soc.religion.christian. **NG5** is from: comp.os.ms-windows.misc, misc.forsale, rec.sport.hockey, sci.space, soc.religion.christian. **NG6** is from: comp.graphics, rec.sport.hockey, sci.crypt, sci,space, soc.religion.christian, talk.politics.gun as in [22].

#### 3.1.3 Reuters-21578

Reuters-21578 news collection contains news articles collected from the Reuters newswire in 1987. We create three datasets using 200 documents from each category. **R4** contains documents from crude, earn, grain, money-fx. **R5** contains documents from: coffee, crude, interest, sugar, trade. **R6** contains documents from acq, crude, earn, grain, money-fx and ship as used in [22].

### 3.2 Method

We use a GA to specify a set of search queries in Apache Lucene format. The documents returned by each query is a cluster. We use a simplified example based on the problem of clustering documents in the Newsgroup 5 (**NG5**) dataset to assist the explanation. To begin we assume the simpler case of evolving single word queries. We will then go onto to explain the extra requirements needed when building multiword queries.

*Step 1*: pre-processing.

Before we start evolving queries, all the text is placed in lower case and a small stop set is used to remove common words with little semantic weight. For each dataset, an inverted index is constructed from the collection of documents so that for each term in the collection the list of documents where the term occurs is recorded. Each document is also labelled according to its pre-set category. Of course, the GA has no access to the category label which is only used to evaluate the effectiveness of the clustering once all the stages have completed.

*Step 2*: create a wordlist.

In the second step, we create an ordered list of significant words (terms) which is used by the GA for building queries. To construct the list, the TF*IDF (term frequency * inverse document frequency) value for each term in the collection is calculated. TF is the number of occurrences of a term in a document and IDF is the inverse of the number of documents in which the term occurs. TF*IDF (often used in term weighting) is used to identify terms that are concentrated in particular documents and may therefore be of more significance in a collection. For each term in the index, we determine TF*IDF values occurring in each document as indicated below, where *terms* is the set of terms and *documents* is the set of documents in the index. We have modified the basic TF*IDF calculation to give extra weighting to uncommon words.

For Documents *D*, Terms *T*. For $t \epsilon T, d \epsilon D$ if *t* occurs in document *d* and write #(t, d) for the number of occurrences of *t* in *d*

$$DF(t) = |\{d \in D : t \in d\}|$$

$$IDF(t) = log\left(1 + \frac{|D|}{DF(t)}\right)$$

$$TFIDF(t) = \sum_{d \in D} \left(\sqrt{\#(t, d) * (2 - IDF(t))}\right)$$

The value for each term is computed and the list of terms is sorted by this value. The top 100 words are selected from the list for use in GA query building. This step is only required once for each index and is calculated before the start of the evolution, after which the list is fixed. The index is simply the words place in the TF*IDF ordering. In the example shown in Table 1 the length of the list is only 8.

*Step 3*: Create generation 0.

Table 2 shows a sample chromosome from the population of generation 0. Chromosomes have an integer representation where the values can be in the range [0.. 100] (the maximum size of the wordlist).

*Step 4*: determine *k* (the number of categories).

If *k* is predefined, then this step can be omitted. In the example in Table 2 *k* is genome defined as 5 An int value in the inclusive range [2.. 9] (8 possible cluster sizes) is used.

*Step 5*: build a set of *k* queries.

In the example shown in Table 2 each gene defines a single word search query (SQ) and each search query defines the cluster as the set of documents which contain that word.

*Step 6*: fire each query in the set.

In our example, five single word search queries are generated for the NG5 dataset. For each individual in the population, fire each of the search queries and determine its fitness by examining the clusters of documents returned by the queries and counting the total number of documents returned which occur in only 1 cluster (see fitness calculation below).

*Step 7*: repeat.

Repeat steps 4–6 for 100 generations (termination criteria) and select the individual with the highest fitness.

*Step 8*: apply genetic operators.

Apply genetic operators to create a new generation.

*Step 9*: create initial document clusters.

The selected individual at the end of a run will produce a set of single word search queries. Fire each of these queries and save the document clusters produced for the KNN stage. Remove any documents which are returned by more than 1 query.

*Step 10*: KNN.

Some documents in the collection may not contain any of the query words or are returned by more than one query and are therefore not included in any of the initial clusters produced by the GA. We use the K-Nearest Neighbour (KNN) algorithm to add any unassigned document to its closest cluster (see Sect. 3.7 below for a detailed explanation).

**Table 1** Word list

| 0 | 1 | 2 | 3 | 4 | 5 | 6 | 7 |
|---|---|---|---|---|---|---|---|
| Space | Nasa | God | Orbit | Hockey | File | Sale | Game |

**Table 2** Creating single word search queries

|  | k | SQ0 | SQ1 | SQ2 | SQ3 | SQ4 |
|---|---|---|---|---|---|---|
| Chromosome: | 5 | 0 | 4 | 5 | 1 | 7 |
| Query words: |  | Space | Hockey | File | Nasa | Game |

**Fig. 1** eSQ steps

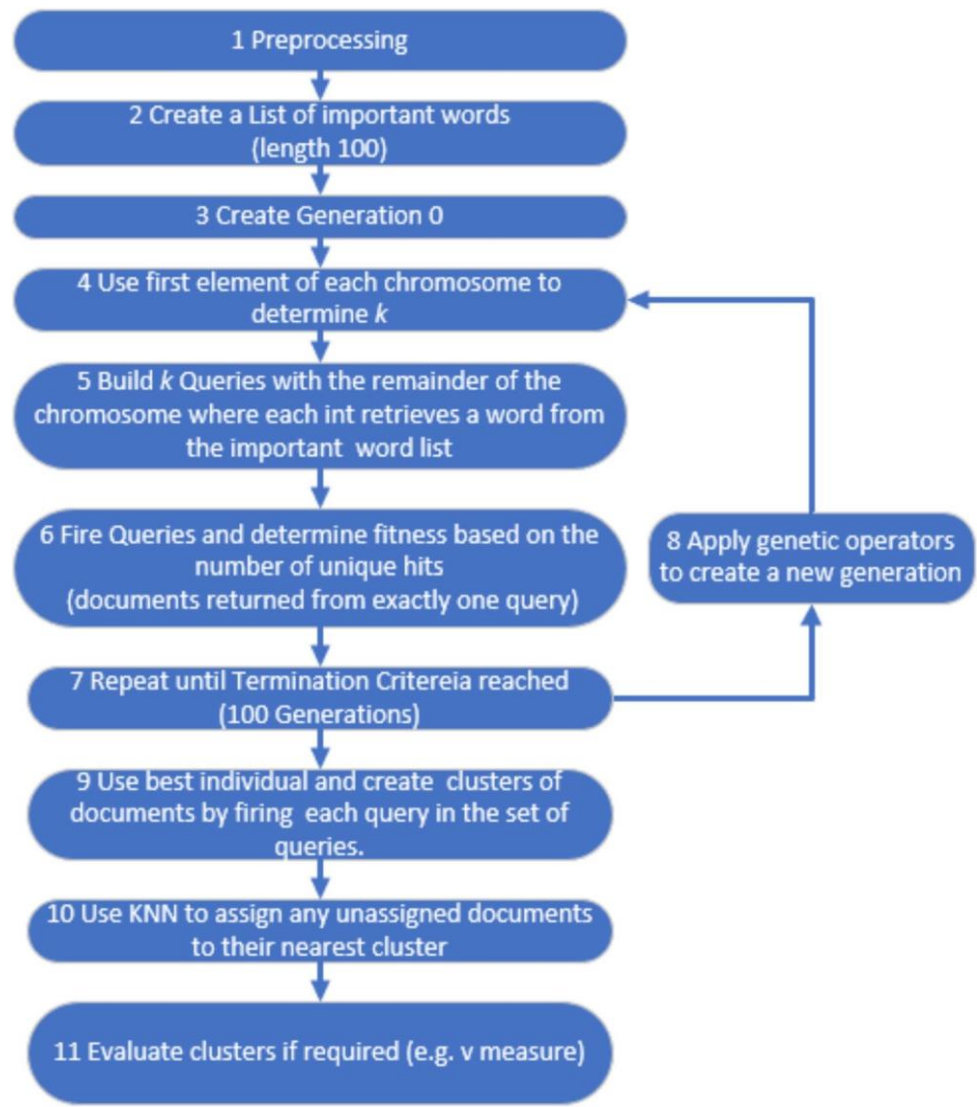

**Table 3** GA parameters

| Parameter | Value |
| --- | --- |
| Selection type | Tournament |
| Subpopulations | 4 |
| Population size | 512 |
| Generations | 100 |
| Crossover probability | 0.8 |
| Mutation probability | 0.1 |
| Elitism | Best 2 |

*Step 11*: evaluation.

If evaluation is required, measure the final V Measure and Adjuster Rand Index value of the expanded clusters with reference to the original category labels.

The steps are summarised in Fig. 1

A GA contains many random elements, so we therefore repeat each run 11 times.

### 3.2.1 Parameters

We used a fixed set of standard GA parameters in all our experiments which are summarised in Table 3. We use an island model with 4 subpopulations as a means to increase diversity and exchange 3 individuals every 30 generations.

### 3.2.2 Multi-word queries

We can build multiword queries by extending the length of the genome, for example doubling the length of the genome to allow for two-word queries and taking the modulus of $k$ to determine which query each gene relates to. A word can only be added once to a set of queries: if the genome specifies two or more occurrences of a particular word, only the first occurrence is used. Where a query is made of two or more words they are connected with a logical OR (disjunction) such that documents are returned which contain any of the words in the query.

When building a query specified by a chromosome, we have found it useful to add a requirement for queries made of two or more words. Each word in a multi-word query can also be used as a single word query. Before we add a new word (*newWord*) to a query already containing a word (*rootWord*), we must first check that the intersect requirement is met by calculating the following:

| | |
|---|---|
| andCount: | count of documents containing the newWord AND the rootWord |
| newWordCount: | count of documents containing the newWord |
| intersectRatio: | andCount/newWordCount |

We have experimented with various values for the minimum *intersectRatio* and have found 0.5 to be a suitable value (see results section below). If *intersect Ratio* > = 0.5 the word is added to the query otherwise nothing is added. To put it another way, before we add a new word to a query, we check that at least 50% the documents which contain the new word also contain the root word. This method also has the advantage of making the first word in a query more likely to be a good cluster label.

### 3.3 Example generating 3 search queries (SQ0, SQ1, SQ2)

Table 4 shows and example where *k* is determined in the first gene of the chromosome and the rest of the chromosome is used to build up a multi-word query.

In this case the chromosome specifies a *k* value of 3 meaning that 3 clusters will be created. The 3 queries shown in Table 5 will be created.

### 3.4 Fitness calculation

Text clustering aims to return sets of documents which are related to each other but not related to documents in other clusters. We have created and tested two fitness functions that aim to partition a document collection into clusters by generating a set of search queries. The first fitness function is for the case where the desired number of clusters (*k*) is known in advance. In the second case, the GA will attempt to determine the optimal value for *k*.

When calculating fitness from a set of queries generated by a chromosome, we define *uniqueHits* as the count of documents returned by exactly one query in the set of queries.

Let $Q$ be a set of queries, let $D$ be a set of documents. Let $M \subseteq Q \times D$ be the set of pairs *(q, d)* where query $q \in Q$ matches document $d \in D$:

*uniqueHits*: $|\{d \in D : \exists ! q \in Q, (q,d) \in M\}|$

For the case where *k* is known in advance, we have found that *uniqueHits* is a good fitness measure where the higher the value (the number of documents returned by exactly one query) the better the fitness.

We have noticed that in the case where *k* is defined in the chromosome the GA often produces solutions with too many categories with respect to the labelled collections. In fact, this is to be expected since overlapping clusters do not lead to a reduction in fitness. We found that introducing a small penalty for more clusters, as in the second fitness test (below), improved effectiveness.

*uniqueHits* * (1 - (*k* * penalty))

We have found a suitable value for penalty to be 0.02. This value is examined in the results section. In algorithm 2 below we show pseudo code to calculate *uniqueHits*.

**Table 4** Chromosome to determine *k* and create 3 search queries (SQ)

| Representation | K | SQ0 | SQ1 | SQ2 | SQ0 | SQ1 | SQ2 | SQ0 | SQ1 | SQ2 |
|---|---|---|---|---|---|---|---|---|---|---|
| Chromosome | 3 | 0 | 7 | 2 | 3 | 3 | 5 | 1 | 4 | 2 |

**Table 5** Creating multi-word queries

| | Gene | Specified words | Final query | Comment |
|---|---|---|---|---|
| SQ0 | 0,3,1 | *Space, orbit, nasa* | *Space* OR *orbit* OR *nasa* | *Orbit* and *nasa* both have a high intersect ratio with root word *space* |
| SQ1 | 7,3,4 | *Game, orbit, hockey* | *Game* OR *hockey* | *Orbit* does not meet the intersect requirement for the root word *game* so is not included in the final query |
| SQ2 | 2,5,2 | *God, file, god* | *God* | *File* does not meet the intersect requirement for the root word *god*. Repeated word is ignored |

```
int countUniqueHits(querySet, D)
{
  int uniqueHits = 0
  for each d in D
  {
      if there is q in querySet where Match(q, d)
      but for all q' ≠ q in querySet NOT Match(q', d)
      then uniqueHits++
  }
  return uniqueHits
}
```

**Algorithm 2**

querySet is the set of queries (of size *k*) which have been generated by a single chromosome in the population. The value of *uniqueHits* will be the number of documents in the collection which are returned by exactly one query in the querySet.

### 3.5 Effectiveness measures

Effectiveness is determined by referring to the original category labels (ground truth) from the relevant collection.

#### 3.5.1 V measure

We use V-measure [38] as the primary method of assigning effectiveness. The V-measure is based on a combination of homogeneity ($h$) and completeness ($c$). A perfectly homogeneous clustering is one where each cluster has data-points belonging to the same class label. Homogeneity describes the closeness of the clustering algorithm to this perfection.

A perfectly complete clustering is one where all data-points belonging to the same class are clustered into the same cluster. Completeness describes the closeness of the clustering to this perfection. The V-measure score is the harmonic mean of homogeneity and completeness as given by

$$V = \frac{(1 + \beta) * h * c}{(\beta * h + c)}$$

We assign a default value of 1 to beta so that homogeneity and completeness are given equal weighting.

#### 3.5.2 Adjusted rand index

We also provide the adjusted Rand Index (ARI) [12] as a secondary performance measure. ARI is a measure of the similarity between the clusters produced by the algorithm and the original document labels. The ARI is calculated as follows:

$$ARI = \frac{2(agreement - chance)}{agreement + chance}$$

where:

- Agreement is the number of pairs of points that are assigned to the same cluster in both clusters.
- Chance is the expected number of pairs of points that would be assigned to the same cluster by chance, given the number of clusters and the size of the data set.

The ARI can take values between -1 and 1, where -1 indicates perfect disagreement and 1 indicates perfect agreement. A value of 0 indicates that the two clusters are no better than random.

#### 3.5.3 Cluster count error

We also provide the cluster count error which is simply the absolute value of the number of classes minus the number of clusters. This measure is only relevant for the case where *k* is not known in advance.

### 3.6 Definitions

In this section we provide a more formal definition of the query-based clustering and link this with the V measure described above.

Let *W* be the set of all words in any document in a collection, so $W \subseteq \wp(\Sigma^*)$, here $\wp(X)$ is the power set of *X*, $\Sigma$ is a finite alphabet and $\Sigma^*$ is the set of finite strings over $\Sigma$.

We consider a document as an unstructured set of words, i.e. a document belongs to $\wp(W)$. In this way, we ignore the order and multiplicity of the words in the document. Later, we may consider a document as a multi set of words. Let $D \subseteq \wp(\wp(W))$ be a set of documents.

In the simplest case, a query is a single word $w \in W$. By membership, each query defines the set $\delta(w) \subseteq D$, of all documents d such that $w \in d$. More generally, a query $q$ is a set of words (i.e. $q \in \wp(W)$). The query $q$ matches the document $d \in D$ if and only if $q \cap d \neq \emptyset$ i.e. if at least one word of the query occurs at least once in the document. Again, for any query $q$ we define $\delta^f(q) \subseteq D$ to be the set of documents d such that $q$ matches $d$. Observe that

$$\delta^f(q) = \bigcup_{w \in q} \delta(w) \tag{1}$$

A chromosome $(q_i : i < k)$ is a sequence of $k$ queries (some $k > 1$). A *uniqueHit* for a chromosome occurs when a document matches exactly one of its $k$ queries. The *uniqueHitcount* for a chromosome is the count of documents matching exactly 1 query in the set. Let

$$u(q_i : i < k) = \bigcup_{i<k} \left( \delta^f\, q_i \setminus \bigcup_{j \neq i, j < k} \delta^f\, q_j \right) \tag{2}$$

The symmetric difference of the $\delta\, q_i\, s$, i.e., the set of all documents that are matched by exactly one of the $k$ queries.

A *class labelling* of $D$ is any finite partition of $D$, so a class labelling $\{S_i : i < s\}$ consists of $s$ disjoint non-empty sets (some finite $s$) whose union is the whole of $D$. $s$ is the size of the partition. So, a class labelling belongs to $\wp(\wp(D))$.

Let $C = \{S_i : i < s\}$ be a class labelling of size $s$, and let $K = \{q_j : j < k\}$ be a set of $k$ queries for some finite $k$. For $j < k$ let $Q_j = \delta^f(q_j) \subseteq D$, the set of documents that match $q_j$.

The following definitions, used to compute the V-measure of a set of clusters and a set of categories, are standard, see for example [38] for a fuller motivation and explanation of terms and the measure. We may define the V-measure of $(C, K)$ as shown in Eq. 3.

$$V(C, K) = \frac{2hc}{h + c} \tag{3}$$

where

$$h = 1 - \frac{H(C|K)}{H(C)}$$

$$c = 1 - \frac{H(K|C)}{H(K)}$$

$$H(C|K) = - \sum_{i<s, j<k} \frac{|S_i \cap Q_j|}{D} . \log \frac{|S_i \cap Q_j|}{Q_j}$$

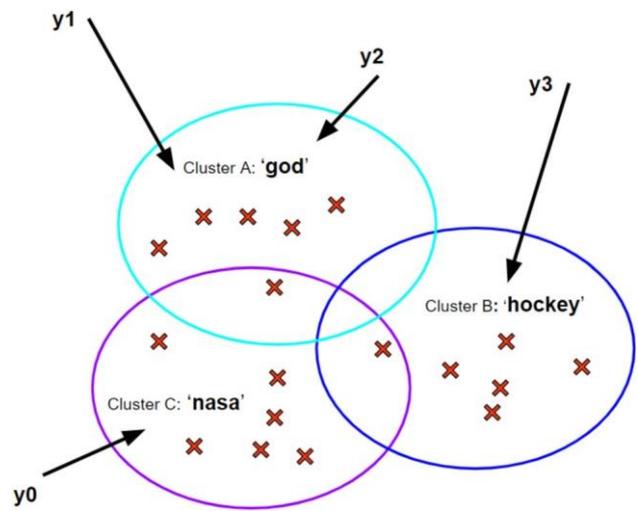

Fig. 2 KNN expansion

Table 6 Clusters generated with no intersect requirement

| Cluster | Query words | Document hits |
|---|---|---|
| 1 | Hockey nhl game players | 208 |
| 2 | Sale please mail windows god space high work apr anyone | 1357 |

$$H(K|C) = - \sum_{i<s, j<k} \frac{|S_i \cap Q_j|}{D} . \log \frac{|S_i \cap Q_j|}{S_i}$$

$$H(C) = - \sum_{i<s} \frac{\sum_{j<k}|S_i \cap Q_j|}{s} . \log \frac{\sum_{j<k}|S_i \cap Q_j|}{s}$$

$$H(K) = - \sum_{j<k} \frac{\sum_{i<s}|S_i \cap Q_j|}{s} . \log \frac{\sum_{i<s}|S_i \cap Q_j|}{k}$$

### 3.7 Cluster expansion using KNN

The clusters produced by the GA generated search queries have a drawback in that many of the documents are not returned by any query; on average only 70% of the documents are clustered. If we discard the documents which are not in any cluster and then analyse the remaining documents which are clustered with reference to the original class labels the clusters have a high V-measure, mostly above 0.8 and sometimes approaching 1. However, it is usually the case that we need to add every document to a cluster. To achieve this, we include a second stage whereby the query generated clusters are used as labelled training sets for a classifier. We use a KNN classifier to assign each of the unassigned

Table 7 Single word queries (NG5)

|   | Query word | Document hits |
|---|------------|---------------|
| 1 | Space      | 204           |
| 2 | Windows    | 278           |
| 3 | Team       | 176           |
| 4 | Sale       | 192           |
| 5 | God        | 205           |

Table 8 Multi-word queries with intersect requirement (NG5)

|   | Query words | Document hits |
|---|-------------|---------------|
| 1 | Sale | 192 |
| 2 | Windows files | 295 |
| 3 | Game players hockey games | 299 |
| 4 | God christ jesus church | 294 |
| 5 | Space moon nasa | 262 |

## 4 Results and discussion

### 4.1 Intersect requirement

The intersect requirement was developed to support multi word query building for the case where $k$ is not known in advance. In this situation, if the intersect constraint is not included the GA will almost always select a value of 2 for the number of clusters ($k$). For example, a typical clustering for the NG5 set is shown by the set of 2 queries is shown in Table 6.

The fitness test is based on the count of documents returned by exactly one query, so the query set shown achieves a high fitness, but the number of clusters (2) does not match the number of labelled classes (5), and the second query is returning documents from multiple classes. Completeness is high (0.852) but homogeneity low (0.212) and the V-measure for this clustering is also low (0.340).

We can improve things by restricting the GA to using one word per cluster. In this case, using more queries can result in more unique hits and higher fitness. A typical result is shown with the set of queries below:

The correct number of categories has been identified, and the evaluation metrics show a distinct improvement (v: 0.773, h: 0.773, c: 0.774) (Table 7).

The intersect constraint allows an individual to add more words to a single term query, but only when the set of documents retrieved by the first term (root term) in a query intersects with the set of documents retrieved by any new term added to the query. In the results shown below we require that 50% of the documents retrieved by the new term are documents to their nearest cluster. Figure 2 shows the NG3 collection where an X represents a document which has been assigned to a category. For example, cluster A shows all the documents which contain the word 'god'. Y indicates a document which does not contain any of the search query words ('god', 'hockey' or 'nasa') and are therefore not included in any cluster. The arrows represent the process whereby the Lucene implementation of KNN assigns documents y0–y3 to the nearest cluster. We use a Euclidean distance measure with a K value of 10.

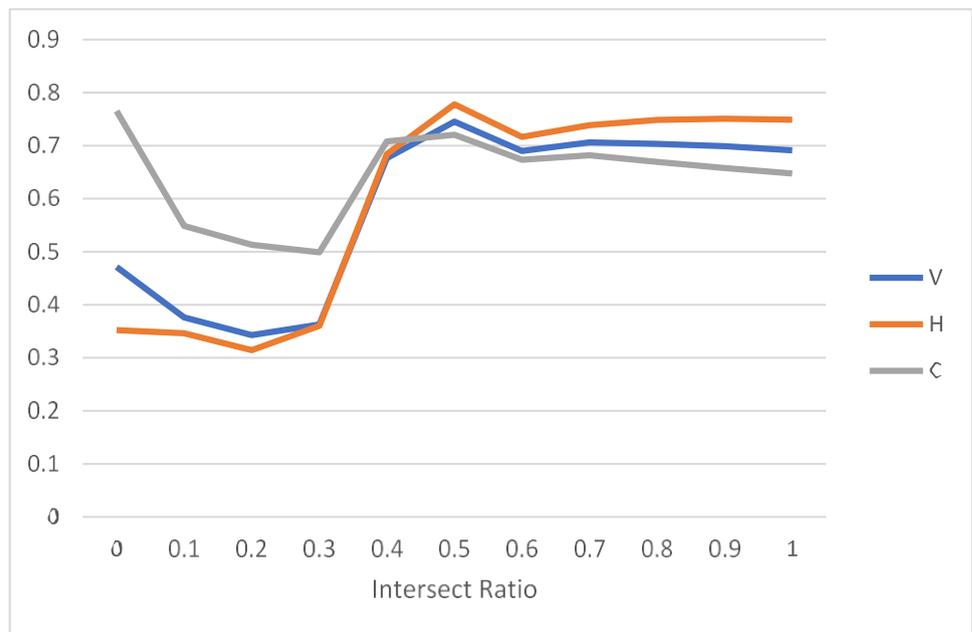

**Fig. 3** Average V, H and C scores across all 8 datasets for different values of the minimum intersect ratio where k is discovered, and multi-word queries are enabled

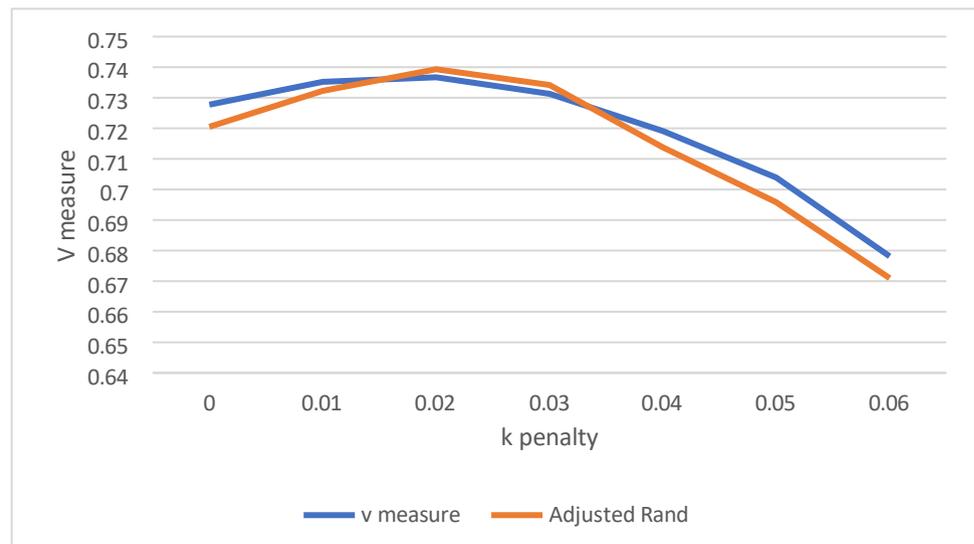

**Fig. 4** Average values for V and Adjusted Rand Index across all 8 datasets for different values of k penalty where k is discovered, and multi-word queries are enabled

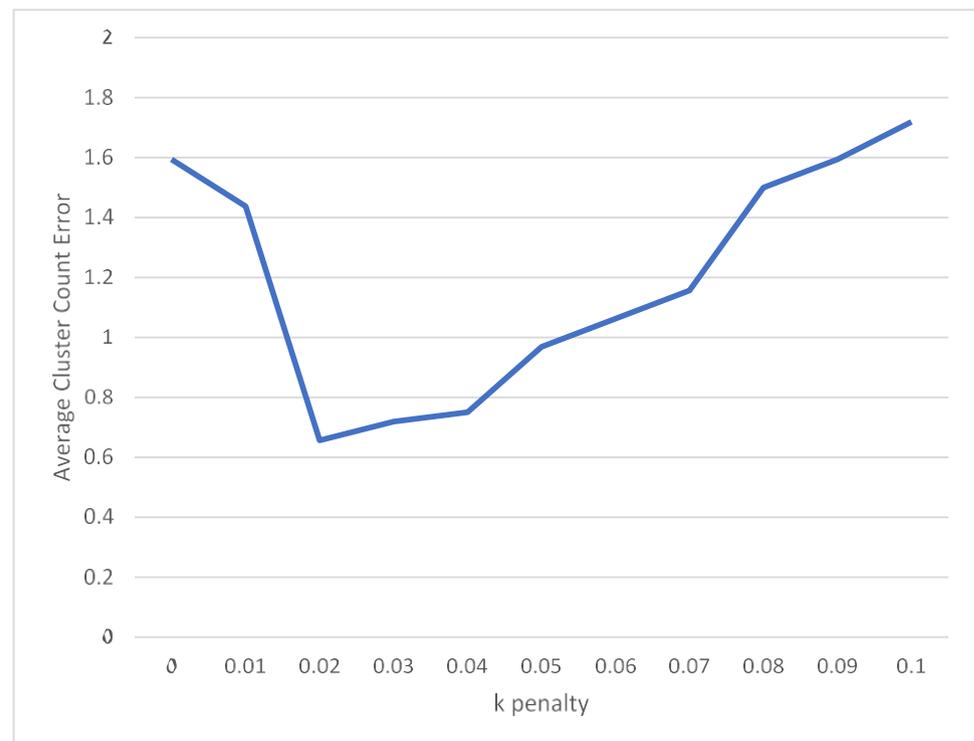

**Fig. 5** Average cluster count error across all 8 datasets for different values of k penalty

found in the set retrieved by the root term. The rationale for the intersect requirement is to create a mechanism which allows GAs to produce queries with multiple terms, but only retrieving related documents ideally from a single category. Using the intersect requirement we see an even bigger improvement. A typical result of a run using the intersect constraint is shown in Table 8.

The GA can add more keywords to queries provided the intersect requirement is met for each new term. The set of queries above has correctly created 5 clusters with (v: 0.882, h: 0.880, c: 0.883).

We ran the GA across all the indexes with various values between 0 and 1 for the intersect requirements and present the results in Fig. 3.

Following these results, we use an intersect ratio of 0.5 in the experiments described below.

### 4.2 Penalty for more clusters

In the situation where $k$ is not known in advance, the number of clusters produced by the GA is typically higher than the number of categories existing in the original collection. This

**Fig. 6** v measure for KNN expansion

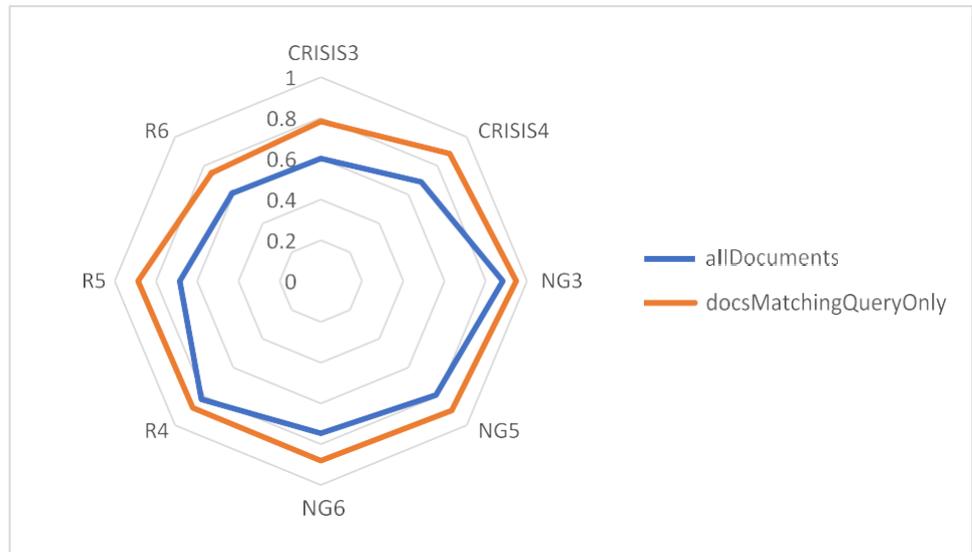

**Fig. 7** Results overview

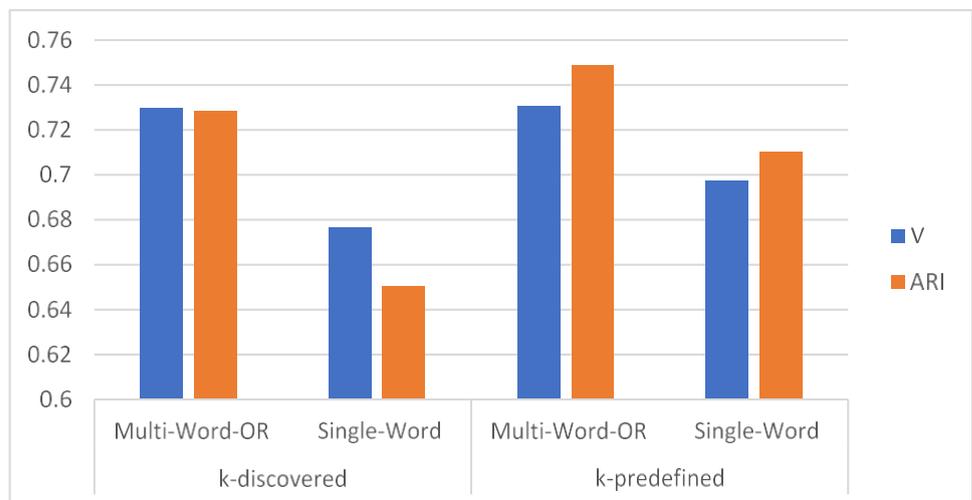

higher fragmentation leads to weaker results, and we found effectiveness could be improved by introducing in the fitness a small penalty based on the number of clusters:

Fitness = uniqueHits ∗ (1.0 − (kPenalty ∗ k))

We investigated various values for the penalty (k penalty) as shown in Fig. 4

We also investigated how the cluster count error responds to different values of $k$ penalty as shown in Fig. 5.

V measure and Adjusted Rand peak with a $k$ penalty of 0.02 and the cluster count error is at its lowest for this value. Following these results, we selected a penalty of 0.02 in the experiments described below for the case where $k$ is not known in advance.

### 4.3 KNN expansion

There may be cases where it is useful to return more accurate clusters by excluding the documents where we are less confident of cluster membership. Figure 6 shows the effect of cluster expansion on the v measure. For all indexes the v measure is reduced after the expansion. This is to be expected as we are now including all documents in the collection rather than only the documents returned by at least one query in the set of queries. The average v measure where we use only documents matching a query is 8.64.

### 4.4 Overview

Multi-word queries perform better than single word queries. Where $k$ is given in advance results are slightly improved (Fig. 7).

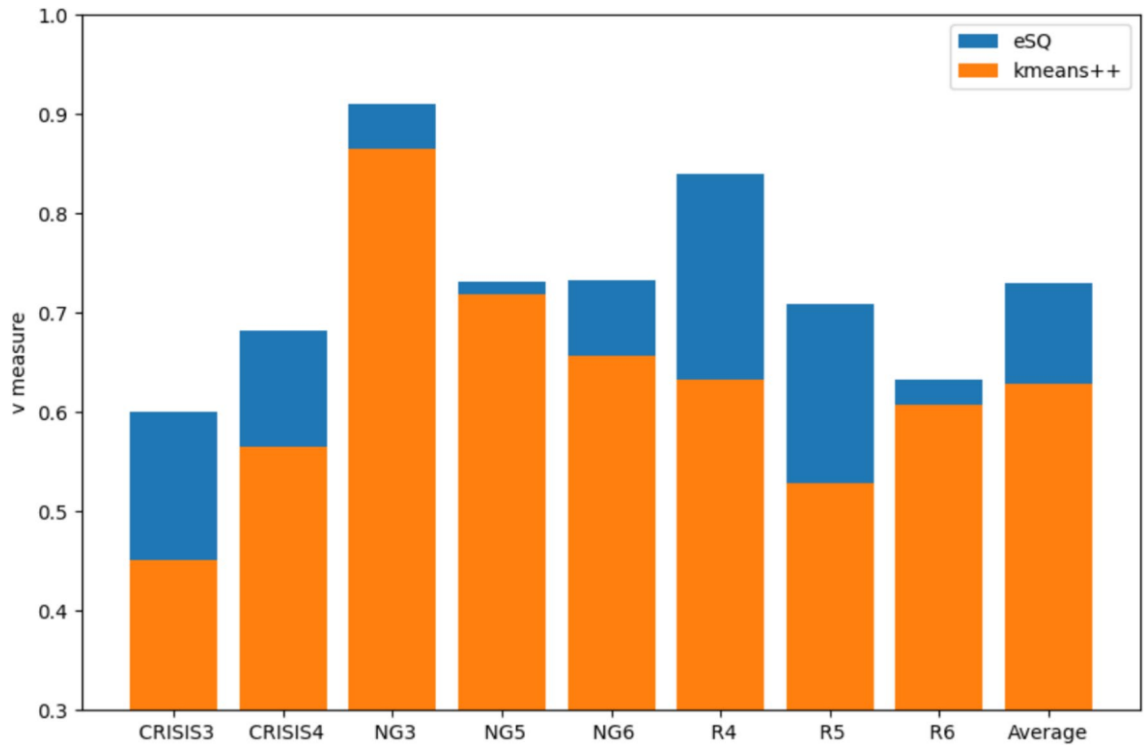

**Fig. 8** v measure for eSQ (k discovered) and k-means++

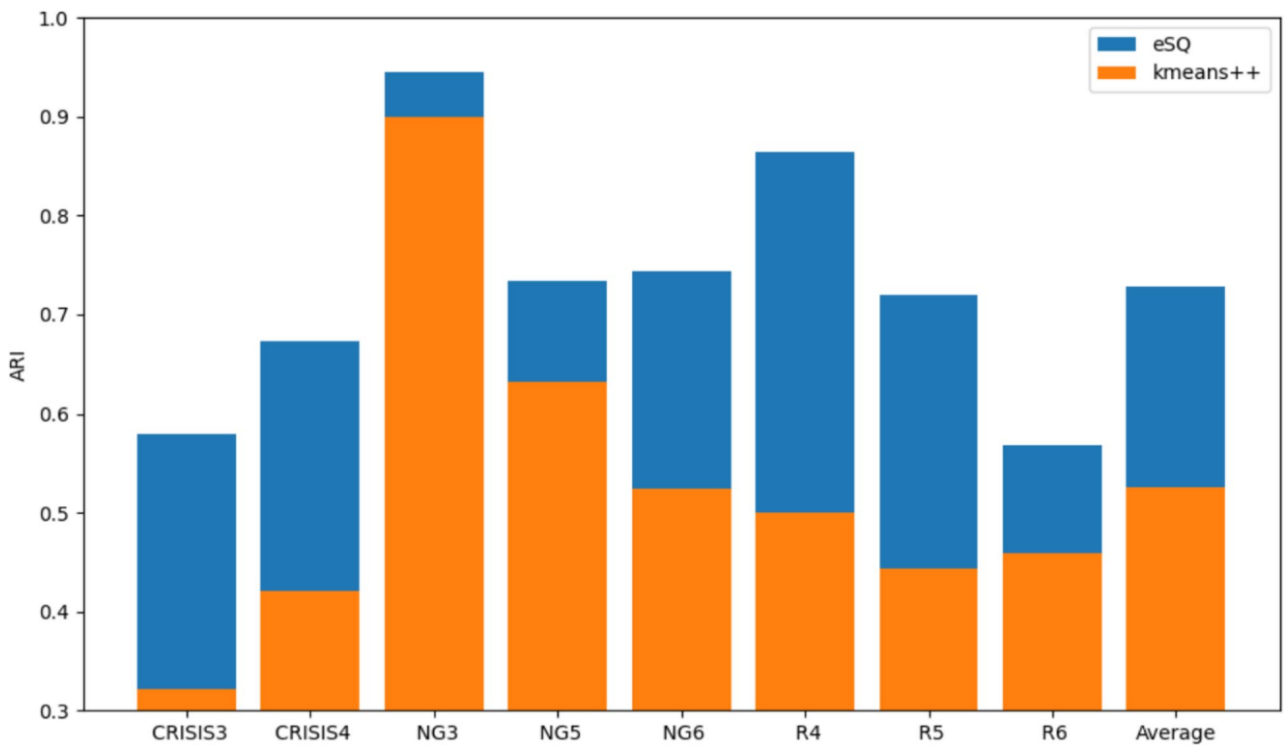

**Fig. 9** ARI for eSQ (k discovered) and k-means++

**Table 9** Average across all indexes

|  | Query type | v-measure | ARI |
|---|---|---|---|
| k-discovered | Multi-word | 0.730 | 0.728 |
|  | Single-word | 0.677 | 0.650 |
| k-predefined | Multi-word | 0.731 | 0.749 |
|  | Single-word | 0.697 | 0.710 |

**Table 10** Comparison of eSQ and k-means + +

|  | v | | ARI | |
|---|---|---|---|---|
|  | eSQ | k-means + + | eSQ | k-means + + |
| Crisis3 | 0.600 | 0.451 | 0.580 | 0.322 |
| Crisis4 | 0.682 | 0.565 | 0.673 | 0.421 |
| NG3 | 0.910 | 0.865 | 0.944 | 0.9 |
| NG5 | 0.731 | 0.718 | 0.734 | 0.632 |
| NG6 | 0.733 | 0.657 | 0.744 | 0.524 |
| R4 | 0.841 | 0.632 | 0.864 | 0.5 |
| R5 | 0.709 | 0.529 | 0.720 | 0.444 |
| R6 | 0.633 | 0.607 | 0.568 | 0.459 |
| Average | 0.730 | 0.628 | 0.728 | 0.525 |

**Table 11** Standard deviation of V and ARI

|  | v | | ARI | |
|---|---|---|---|---|
|  | eSQ | k-means + + | eSQ | k-means + + |
| Crisis3 | 0.007 | 0.121 | 0.008 | 0.157 |
| Crisis4 | 0.005 | 0.081 | 0.005 | 0.106 |
| NG3 | 0.002 | 0.025 | 0.002 | 0.030 |
| NG5 | 0.009 | 0.070 | 0.009 | 0.116 |
| NG6 | 0.005 | 0.041 | 0.005 | 0.069 |
| R4 | 0.028 | 0.071 | 0.035 | 0.126 |
| R5 | 0.012 | 0.062 | 0.015 | 0.071 |
| R6 | 0.029 | 0.081 | 0.040 | 0.125 |
| Average | 0.099 | 0.138 | 0.109 | 0.194 |

**Table 12** Time in milliseconds

|  | eSQ | k-means + + |
|---|---|---|
| Crisis3 | 524 | 15 |
| Crisis4 | 717 | 20 |
| NG3 | 523 | 18 |
| NG5 | 1024 | 39 |
| NG6 | 1440 | 51 |
| R4 | 703 | 18 |
| R5 | 932 | 19 |
| R6 | 1220 | 25 |
| Average | 885 | 26 |

**Table 13** v measure for agglomerative and spectral clustering

|  | Agglomerative | Spectral |
|---|---|---|
| Crisis3 | 0.243 | 0.040 |
| Crisis4 | 0.293 | 0.067 |
| NG3 | 0.309 | 0.813 |
| NG5 | 0.364 | 0.610 |
| NG6 | 0.403 | 0.639 |
| R4 | 0.446 | 0.433 |
| R5 | 0.474 | 0.302 |
| R6 | 0.467 | 0.326 |
| Average | 0.375 | 0.404 |

## 4.5 Comparison with k-means + +

We present a basic comparison across the 8 datasets for the eSQ (multi-word) system and the implementation of k-means + + in scikit-learn [12]. The value of $k$ is given in advance for k-means + + but we show the results for eSQ where $k$ is discovered (a harder problem). 11 runs were obtained for both systems and the average value of the V measure and ARI is shown. We use a tf-idf based vectorizer and a feature size of 1000 for k-means + + .

These results are visualized in Fig. 8 and Fig. 9 showing that eSQ ($k$ discovered) outperforms k-means + + in every dataset (Tables 9 and 10).

Table 11 compares the standard deviation of the results across the 11 runs.

Table 12 show the time in milliseconds to achieve the clustering for each index. All programs were run on an Intel i7-10,700 CPU running at 2.90GHz. GAs are known as a resource intensive approach and the eSQ system is significantly slower when compared to k-means + + . We should note that although we have tried to optimize the eSQ code this has not been the focus of the development, and we believe there is plenty of room for improvement (Table 13).

## 4.6 Comparison with agglomerative clustering and spectral clustering

We also applied agglomerative clustering and spectral clustering [39], using the implementation available in scikit learn [12] with tf-idf vectorization. Agglomerative clustering also has the advantage of not requiring the number of clusters to be provided in advance, however both methods performed quite poorly compared to k-means + + or eSQ.

Spectral clustering is often performing better than agglomerative clustering but is failing to effectively cluster the short text (tweet) data in the crisis datasets. Our findings suggest that agglomerative clustering struggled due to the inter-relation of documents between the classes. For example, in the NG6 dataset, the graphic class comprises documents related to graphics, programming, and computing. The agglomerative (bottom-up) clustering algorithm is somewhat rigid as once two data points are joined together to form a cluster, they may not rejoin another cluster at a later stage. This also contributed to the performance of spectral clustering because the algorithm attempts to model the local neighbourhood relationships between the data points [40].

## 5 Conclusions and future work

We have presented eSQ, a novel system for text clustering which is based on a set of GA generated search queries. The system takes a hybrid approach whereby the GA operates in an unsupervised manner to produce initial clusters from the documents returned by each query in the set. Unlike existing clustering systems this step does not require us to compute a similarity measure between documents. A second supervised step is then taken where KNN uses the GA clusters as training documents so that each document is assigned to its nearest cluster.

eSQ is different to most modern clustering systems which would use a document or a point in a multi-dimensional space as the cluster centre. The eSQ system can produce effective text clustering by using a search query at the cluster centre and works well even where the number of clusters is not known in advance. The search query method provides an explanation of cluster construction where the search terms can function as cluster labels and also provides the possibility of manually modifying the simple search queries where required. As mentioned in the introduction the cluster hypothesis suggests that search query method will naturally align with information retrieval requirements.

### 5.1 Limitations

We have used a variety of datasets in our development and testing, but we cannot be sure that the results obtained here are generalizable to other types of text data. Furthermore, the datasets we have used have a maximum number of 8 categories and we have not tested the system where a large number of clusters would be required for a good solution.

### 5.2 Recommendations for future study

We are hoping to investigate how supervised-weighting schemes might be used to improve the clustering, for example in the creation of the word list used by the GA. Currently the algorithm only generates disjunctive queries and is only suitable for clustering text documents. We would like to experiment with more complex queries which include conjunction, negation and other search query types that have successfully been used in text classification. We are investigating the possibility of applying the algorithm to cluster other media such as images. Lastly, we will investigate how the eSQ system could be combined with existing techniques for topic modelling.